\documentclass[aps,twocolumn,preprintnumbers,amsmath,amssymb,nofootinbib,superscriptaddress,notitlepage]{revtex4}
\usepackage{graphicx,color,dcolumn,booktabs,bm}
\usepackage{longtable,lscape}
\usepackage{txfonts}
\usepackage{overpic}
\usepackage{amssymb}
\usepackage{amsmath}
\usepackage{epstopdf}
\usepackage{indentfirst}
\usepackage{feynmf}   
\usepackage{slashed}  
\usepackage{cases}
\usepackage{color}
\usepackage{float}
\usepackage{wasysym}
\usepackage{multirow}
\usepackage{ulem}
\usepackage{soul}
\usepackage{times}
\usepackage{cancel}
\usepackage[colorlinks,
            citecolor=blue,
            anchorcolor=red,
            menucolor=red,
            linkcolor=red,
            filecolor=red,
            runcolor=red,
            urlcolor=blue,
            frenchlinks=red]{hyperref}

\graphicspath{{Figures/}} %
\usepackage{CJK}

\begin{document}
\title{$\Lambda_c(2910)$ and $\Lambda_c(2940)$ as the conventional baryons dressed with the $D^*N$ channel}

\author{Zi-Le Zhang}
\affiliation{School of Physical Science and Technology, Lanzhou University, Lanzhou 730000, China}
\affiliation{Research Center for Hadron and CSR Physics, Lanzhou University and Institute of Modern Physics of CAS, Lanzhou 730000, China}

\author{Zhan-Wei Liu}\email{liuzhanwei@lzu.edu.cn}
\affiliation{School of Physical Science and Technology, Lanzhou University, Lanzhou 730000, China}
\affiliation{Research Center for Hadron and CSR Physics, Lanzhou University and Institute of Modern Physics of CAS, Lanzhou 730000, China}
\affiliation{Lanzhou Center for Theoretical Physics, Key Laboratory of Theoretical Physics of Gansu Province, and Frontiers Science Center for Rare Isotopes, Lanzhou University, Lanzhou 730000, China}

\author{Si-Qiang Luo}\email{luosq15@lzu.edu.cn}
\affiliation{School of Physical Science and Technology, Lanzhou University, Lanzhou 730000, China}
\affiliation{Research Center for Hadron and CSR Physics, Lanzhou University and Institute of Modern Physics of CAS, Lanzhou 730000, China}
\affiliation{Lanzhou Center for Theoretical Physics, Key Laboratory of Theoretical Physics of Gansu Province, and Frontiers Science Center for Rare Isotopes, Lanzhou University, Lanzhou 730000, China}

\author{Fu-Lai Wang}\email{wangfl2016@lzu.edu.cn}
\affiliation{School of Physical Science and Technology, Lanzhou University, Lanzhou 730000, China}
\affiliation{Research Center for Hadron and CSR Physics, Lanzhou University and Institute of Modern Physics of CAS, Lanzhou 730000, China}
\affiliation{Lanzhou Center for Theoretical Physics, Key Laboratory of Theoretical Physics of Gansu Province, and Frontiers Science Center for Rare Isotopes, Lanzhou University, Lanzhou 730000, China}

\author{Bo Wang}\email{wangbo@hbu.edu.cn}
\affiliation{School of Physical Science and Technology, Hebei University, Baoding 071002, China}
\affiliation{Key Laboratory of High-precision Computation and Application of Quantum Field Theory of Hebei Province, Baoding 071002, China}
\affiliation{Research Center for Computational Physics of Hebei Province, Baoding, 071002, China}

\author{Hao Xu}\email{xuh2020@nwnu.edu.cn}
\affiliation{Institute of Theoretical Physics, College of Physics and Electronic Engineering, Northwest Normal University, Lanzhou 730070, China}
\affiliation{Lanzhou Center for Theoretical Physics, Lanzhou University, Lanzhou 730000, China}

\begin{abstract}
In this work, we treat $\Lambda_c(2910)^+$ and $\Lambda_c(2940)^+$ as the conventional $udc$ cores dressed with the $D^*N$ channel. We provide a possible interpretation to both $\Lambda_c(2910)^+$ and $\Lambda_c(2940)^+$ within the same framework. In the study, we consider not only the effects between the conventional triquark core and the $D^*N$ channel but also the $D^*N$-$D^*N$ interactions. The mass of $\Lambda_c$ state with $J^P=1/2^-$ is larger than that with $J^P=3/2^-$ in this unquenched picture, which is very different from the prediction of the conventional quenched quark model. Based on the mass spectrum, the spin-parity of $\Lambda_c(2940)^+$ is more likely to be $1/2^-$ while $\Lambda_c(2910)^+$ prefers $3/2^-$. We look forward to the future experiments can test our results with more precise experimental data.
\end{abstract}

\maketitle

\section{Introduction}\label{sec1}
Until now, a growing number of charmed baryons have been reported with the accumulation of experimental data~\cite{ParticleDataGroup:2022pth}. Many of them can properly fit into the conventional charmed baryon spectrum~\cite{Ebert:2011kk,Garcilazo:2007eh,Chen:2014nyo,Capstick:1986ter,Yu:2022ymb,Cheng:2021qpd,Kim:2021ywp}, such as $\Lambda_c(2286)^+$, $\Lambda_c(2595)^+$, $\Lambda_c(2625)^+$, $\Lambda_c(2760)^+$, $\Lambda_c(2860)^+$, $\Lambda_c(2880)^+$,  $\Sigma_c(2455)^{0,+,++}$, $\Sigma_c(2520)^{0,+,++}$, $\Sigma_c(2880)^{0,+,++}$, and so on. The advanced experimental progress have motivated theorists to explore their properties in many theoretical methods~\cite{Chen:2016spr, Crede:2013kia, Kato:2018ijx, Chen:2016iyi,Chen:2022asf,Chen:2016qju,Liu:2019zoy}. We believe the studies of the charmed baryons can deepen our understanding of the nonperturbative behavior of the quantum chromodynamics (QCD) in the low energy regions.

Months ago, the Belle Collaboration reported a new structure called $\Lambda_c(2910)^+$, via  the $\bar{B}^0\to \Sigma_c(2455)\pi\bar{p}$ decay process~\cite{Belle:2022hnm}. Its mass and width are measured to be $2913.8\pm5.6\pm3.8$ $\rm{MeV}$ and $51.8\pm20.0\pm18.8$ $\rm{MeV}$, respectively. In fact, the observed $\Lambda_c(2910)^+$ is a continuation of the experimental studies of $\Lambda_c$ baryons in the past. In 2006, the $BABAR$ Collaboration released the observation of $\Lambda_c(2940)^+$ in the $D^0p$ invariant mass spectrum~\cite{BaBar:2006itc}. Later, Belle confirmed $\Lambda_c(2940)^+$ in the decay mode $\Lambda_c(2940)^+\to \Sigma_c(2455)^{0,++}\pi^{+,-}$~\cite{Belle:2006xni}. In 2017, the LHCb collaboration also observed it~\cite{LHCb:2017jym}, and preferred its possible assignment with $J^P=3/2^-$. Till now, Particle Data Group (PDG) \cite{ParticleDataGroup:2022pth} lists its mass and width as $M=2939.6_{-1.5}^{+1.3}$ $\rm{MeV}$ and $\Gamma=20_{-5}^{+6}$ $\rm{MeV}$, respectively. 

In theory, the newly observed $\Lambda_c(2910)^+$ is studied in several works. $\Lambda_c(2910)^+$ was interpreted as $\Lambda_c(2P,1/2^-)$ with the QCD sum rule~\cite{Azizi:2022dpn}. In Ref.~\cite{Wang:2022dmw}, they treated $\Lambda_c(2910)^+$ as a good candidate of $\Lambda_c|J^P=5/2^-, 2\rangle_{\rho}$ by investigating the strong decay of the low-lying $1P$-wave of the $\rho$ mode excitation, however the $\Lambda_c|J^P=3/2^-, 2\rangle_{\rho}$ and $\Lambda_c|J^P=1/2^-, 1\rangle_{\rho}$ assignments cannot be excluded. The authors in Ref.~\cite{Lu:2018utx} arranged the $\Lambda_c(2910)^+$ as the $2P$-wave of the $\lambda$ mode excitation with $J^P=1/2^-$ or $3/2^-$ by examining its strong decay. 

In the past, the charmed baryons $\Lambda_c(2P,1/2^-)$ and $\Lambda_c(2P,3/2^-)$ have already been studied in the quenched quark model ~\cite{Garcilazo:2007eh,Chen:2014nyo,Ebert:2011kk,Capstick:1986ter,Yu:2022ymb,Cheng:2021qpd,Kim:2021ywp}, while the strong decays of $\Lambda_c(2P,1/2^-)$ and $\Lambda_c(2P,3/2^-)$ were also studied in Refs.~\cite{Lu:2018utx,Chen:2007xf,He:2011jp,Zhong:2007gp,Lu:2019rtg,Guo:2019ytq,Gong:2021jkb}. It is hard to explain $\Lambda_c(2940)^+$ with simple bare $udc$ structure, because its experimental mass is usually about 100 MeV smaller than the theoretical expectations~\cite{Chen:2014nyo, Ebert:2011kk, Capstick:1986ter}. Thus, it has stimulated great interest of theorists in studying the inner structure of $\Lambda_c(2940)^+$.

An important fact is that $\Lambda_c(2940)^+$ locates below the $D^*N$ threshold about $6$ $\rm{MeV}$, so several theoretical groups treat $\Lambda_c(2940)^+$ as a $D^*N$ molecular state~\cite{He:2006is,Cheng:2006dk,Dong:2009tg,Dong:2010xv,Dong:2011ys,Romanets:2012hm,Cheng:2015naa,Wang:2015rda,Xie:2015zga,Huang:2016ygf}. For example, within the one-boson-exchange model, $\Lambda_c(2940)^+$ was treated as the $D^*N$ molecular state with $I\left(J^P\right)=0(1/2^+)$ or $0(3/2^-)$~\cite{He:2010zq}. The $D^*N$ bound state was also found in $[D^*N]_{J=3/2}^{I=0}$ channel within other models, such as the QCD sum rule, constituent quark model, and chiral quark model~\cite{Zhang:2014ska,Ortega:2012cx,Ortega:2014eoa,Zhang:2019vqe}. In Refs.~\cite{Wang:2020dhf,Meng:2022ozq}, the chiral effective field theory was applied to the $D^*N$ interaction and two bound solutions were found with a little mass gap for the isospin $I=0$ channels, which shows the $\Lambda_c(2940)^+$ could be either  $\left[D^*N\right]_{J=1/2}^{I=0}$ or $\left[D^*N\right]_{J=3/2}^{I=0}$. Thus, the interaction between $D^*N$ plays a key role in forming this physical state.

\begin{figure}[htbp]
	\centering
	\includegraphics[width=250pt]{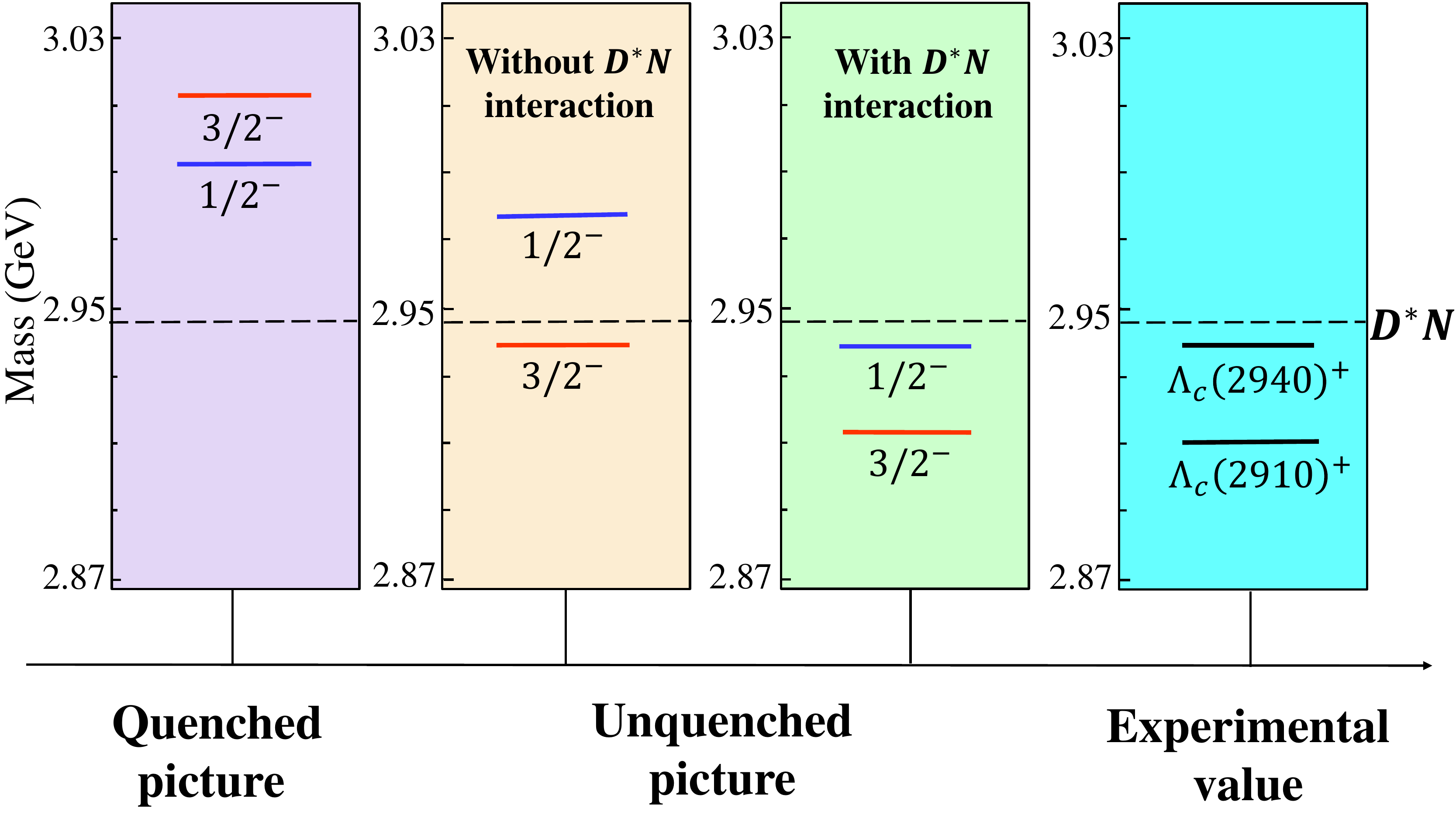}
	\caption{(color online) The masses comparison between the undressed conventional baryons~\cite{Luo:2019qkm}, the conventional baryons dressed by $D^*N$ without considering $D^*N$ interaction~\cite{Luo:2019qkm}, the dressed states with the $D^*N$ interaction in this work, and the experimental baryons \cite{ParticleDataGroup:2022pth,Belle:2022hnm,BaBar:2006itc,Belle:2006xni,LHCb:2017jym}. }\label{fig:1}
\end{figure}

Therefore we conclude that the bare triquark states and the $D^*N$ channel should be equally important for $\Lambda_c(2940)^+$ and $\Lambda_c(2910)^+$. Under an unquenched quark model, the authors in Ref.~\cite{Luo:2019qkm} studied $\Lambda_c(2P,1/2^-)$ and $\Lambda_c(2P,3/2^-)$ considering the coupling to the $D^*N$ channel. They conclude that the mass in spin-$1/2$ becomes larger than that in spin-$3/2$ in the unquenched picture, so that the mass relation is reversed compared to the quenched picture (see Fig. \ref{fig:1}).

 The coupled-channel effects among a bare triquark state and hadron-hadron channels are widely studied in the case of $\Lambda(1405)^0$, $D_{s0}^*(2317)^{\pm}$, $D_{s1}^{*}(2460)^{\pm}$, $X(3872)$, etc~\cite{Heikkila:1983wd,Ono:1983rd,Ono:1985jt,Ono:1985eu,Tornqvist:1984fy,Silvestre-Brac:1991qqx,Pennington:2007xr,Li:2009ad,Barnes:2007xu,Zhou:2011sp,Danilkin:2010cc,Liu:2016wxq,Liu:2016uzk,Liu:2015ktc,Wu:2017qve}. In Refs.~\cite{Zhang:2009bv,Ortega:2012rs,Ortega:2009hj,Li:2009ad}, they studied the structure of $X(3872)$ with coupled $D \bar D^*$ channel, and obtained it as a mixture of bare $c\bar{c}$ core and $D \bar D^*$ component. Moreover, in Refs.~\cite{Luo:2021dvj,vanBeveren:2003kd,Hwang:2004cd,Tan:2021bvl}, the $D_{s0}^*(2317)^{\pm}$ and $D_{s1}^*(2460)^{\pm}$ were also studied by the unquenched quark model and their structure information were revealed as the mixture of bare $c\bar{s}$ cores and $D^{(*)}K$ component. In addition, taking into account the $S$-wave $D^{(*)}K$ interaction, the authors in Ref.~\cite{Yang:2021tvc} studied the $D_s$ states under the Hamiltonian effective field theory, and found the  $D^{(*)}K$ interaction can cause significant mass shifts and lower the mass. According to their investigation, we think including the hadron-hadron interaction is important in the coupled-channel studies.

In the present work, we try to interpret $\Lambda_c(2910)^+$ and $\Lambda_c(2940)^+$ at the same time within a consistent framework with the coupled-channel effects between the bare triquark states and the $D^*N$ channel, as well as the  $D^*N$-$D^*N$ interactions. The physical states $\Lambda_c(2910)^+$ and $\Lambda_c(2940)^+$ can be produced as a competition between the bare $2P$-wave $udc$ cores and $S$-wave $D^*N$ components. We believe our effort can lead us to disclose the true natures of $\Lambda_c(2910)^+$ and $\Lambda_c(2940)^+$.

This article is organized as follows. After the introduction, we present the details of the theoretical framework in Section~\ref{sec2}, which includes the full Hamiltonian and coupled-channel equation involving the interaction of the hadron-hadron channel. The detailed interactions will be given in Section~\ref{sec3} within the chiral effective field theory and quark-pair-creation model. In Section~\ref{sec4}, we present our numerical results and discussion. A short summary follows in Section~\ref{sec5}.

\section{FRAMEWORK} \label{sec2}
If we consider the mixing between the bare state $\left|\Psi_0\right\rangle$ and $|BC,\textbf{p}\rangle$ channel, the physical state~\cite{Weinberg:1965zz,Guo:2017jvc,Tornqvist:1995kr,Kalashnikova:2005ui,Danilkin:2009hr,Eichten:1978tg,Lu:2017hma,Anwar:2018yqm,Ortega:2021fem,Ortega:2021yis,Ortega:2016pgg,Ortega:2009hj} can be represented by
\begin{eqnarray}
\left|\Psi\right\rangle=c_0 \left|\Psi_0\right\rangle+\int {\rm d}^3\textbf{p}~\chi_{BC}(\textbf{p}) \left|BC,\textbf{p} \right\rangle.
\end{eqnarray}
Here, $c_0$ is the possible amplitude to discover the bare state $\left|\Psi_0\right\rangle$ in the physical state $\left|\Psi\right\rangle$, $\chi_{BC}\left(\textbf{p}\right)$ denotes the relative wave function in the hadron-hadron channel $\left|BC,\textbf{p}\right\rangle$, and the normalizing condition is given by
\begin{eqnarray}
    \left|c_0\right|^2+\int \left|\chi_{BC}(\textbf{p})\right|^2{\rm d}^3{\bf p}=1.
\end{eqnarray}
Then the full coupled-channel equation can be formally expressed as
\begin{eqnarray}\label{eq:1}
\left(\begin{array}{cc}\hat{H}_0&\hat{H}_I\\\hat{H}_I&\hat{H}_{BC}\end{array}\right)\left(\begin{array}{c}c_0|\Psi_0\rangle\\\chi_{BC}({\bf p})|BC,{\bf p}\rangle\end{array}\right)=M\left(\begin{array}{c}c_0|\Psi_0\rangle\\\chi_{BC}({\bf p})|BC,{\bf p}\rangle\end{array}\right).
\end{eqnarray}
By expanding Eq.~(\ref{eq:1}), one obtains
\begin{equation}\label{eq:expand1}
\hat{H}_0c_0|\Psi_0\rangle+\hat{H}_I\chi_{BC}({\bf p})|BC,{\bf p}\rangle=Mc_0|\Psi_0\rangle
\end{equation}
and
\begin{equation}\label{eq:expand2}
\hat{H}_Ic_0|\Psi_0\rangle+\hat{H}_{BC}\chi_{BC}({\bf p})|BC,{\bf p}\rangle=M\chi_{BC}({\bf p})|BC,{\bf p}\rangle.
\end{equation}
Here, $\hat{H}_0$ only works on the bare state $|\Psi_0\rangle$, i.e.,
\begin{equation}\label{eq:3}
\hat{H}_0|\Psi_0\rangle=M_0|\Psi_0\rangle,
\end{equation}
where $M_0$ is the bare mass, which can be well determined by the traditional potential models, such as non-relativistic three quark model~\cite{Yoshida:2015tia,Roberts:2007ni,Isgur:1978xj,Isgur:1978wd,Luo:2019qkm}, quark-diquark model~\cite{Chen:2014nyo,Ebert:2011kk,Chen:2016iyi}, Capstick-Isgur model~\cite{Capstick:1986ter}, and so on. For the most low-lying singly charmed baryons, these traditional potential models have good predictive powers in studying the mass spectra. $\hat{H}_0$ is usually written as a sum of two parts, i.e.,
\begin{eqnarray}\label{eq:h0}
	\hat{H}_0=\sum_iE_i+\sum_{i,j}V_{ij},
\end{eqnarray}
where $E_i$ is free energy for $i$-th constituent quark, and the $V_{ij}$ is the effective potential between two quarks where the form depends on specific potential models.

In addition, $\hat{H}_I$ in Eq.~(\ref{eq:1}) represents the transition Hamiltonian between the bare state and the intermediate $BC$ channel. $\hat{H}_{BC}$ is the Hamiltonian describing $BC$-$BC$ interaction. With the above definitions, multiplying $\langle\Psi_0|$ on each side in Eq.~(\ref{eq:expand1}), one obtains
\begin{equation}\label{eq:expand1Mc0}
M_0c_0+\int \chi_{BC}({\bf p})H_{\Psi_0\to BC}^*({\bf p}){\rm d}^3{\bf p}=M c_0.
\end{equation}
Then $c_0$ can be obtained
\begin{equation}\label{eq:c0}
    c_0=\int \frac{\chi_{BC}({\bf p})H_{\Psi_0\to BC}^*({\bf p})}{M-M_0}{\rm d}^3{\bf p}.
\end{equation}
On the other hand, multiplying $\langle BC,{\bf p}^\prime|$ on each side of Eq.~(\ref{eq:expand2}), we have
\begin{equation}\label{eq:expand2chiBC}
    c_0 H_{\Psi_0\to BC}({\bf p}^\prime)+\int \langle BC,{\bf p}^\prime|\hat{H}_{BC}\chi_{BC}({\bf p})|BC,{\bf p}\rangle {\rm d}^3{\bf p}=M\chi_{BC}({\bf p}^\prime).
\end{equation}
The $\hat{H}_{BC}$ is the Hamiltonian of the intermediate $BC$ channel, which includes two parts, i.e., the free and inner interactions. Explicitly, the matrix element sandwiching the $\hat{H}_{BC}$ can be expressed as
\begin{equation}\label{eq:expandHBC}
\begin{split}
&\int \langle BC,{\bf p}^\prime|\hat{H}_{BC}\chi_{BC}({\bf p})|BC,{\bf p}\rangle {\rm d}^3{\bf p}\\
=&\int \langle BC,{\bf p}^\prime|E_{BC}({\bf p})\delta^3({\bf p}-{\bf p}^\prime)\chi_{BC}({\bf p})|BC,{\bf p}\rangle {\rm d}^3{\bf p}\\
&+\int \langle BC,{\bf p}^\prime|V_{BC\to BC}({\bf p},{\bf p}^\prime)\chi_{BC}({\bf p})|BC,{\bf p}\rangle {\rm d}^3{\bf p}\\
=&E_{BC}({\bf p}^\prime)\chi_{BC}({\bf p}^\prime)+\int V_{BC\to BC}({\bf p},{\bf p}^\prime)\chi_{BC}({\bf p}) {\rm d}^3{\bf p},
\end{split}
\end{equation}
where $E_{BC}({\bf p}^\prime)=m_B+m_C+\frac{p^{\prime2}}{2m_B}+\frac{p^{\prime2}}{2m_C}$ is the free energy of the intermediate $BC$ channel and $V_{BC\to BC}({\bf p},{\bf p}^\prime)$ is the hadron-hadron interaction in the momentum space. With Eqs.~(\ref{eq:c0})-(\ref{eq:expandHBC}), we can obtain the following coupled-channel equation
\begin{equation}
\begin{split}
&E_{BC}({\bf p}^\prime)\chi_{BC}({\bf p}^\prime)+H_{\Psi_0\to BC}({\bf p}^\prime)\int \frac{\chi_{BC}({\bf p})H_{\Psi_0\to BC}^*({\bf p})}{M-M_0}{\rm d}^3{\bf p}\\
&+\int \chi_{BC}({\bf p})V_{BC\to BC}({\bf p},{\bf p}^\prime){\rm d}^3{\bf p}=M\chi_{BC}({\bf p}^\prime).
\end{split}
\end{equation}
It is equivalent to the following expression
\begin{equation}\label{eq:in potential}
\begin{split}
\int &\left(\frac{H_{\Psi_0\to BC}^*(\textbf{p})H_{\Psi_0\to BC}(\textbf{p}^{\prime})}{M-M_0}+V_{BC\to BC}(\textbf{p},\textbf{p}^{\prime})\right)\chi_{BC}(\textbf{p})\textrm{d}^3\textbf{p}\\
&+E_{BC}(\textbf{p}^{\prime})\chi_{BC}(\textbf{p}^{\prime})=M\chi_{BC}(\textbf{p}^{\prime}).
\end{split}
\end{equation}
Thus, the final $\chi_{BC}$-coupled-channel equation~(\ref{eq:in potential}) contains the interaction between hadron $B$ and $C$. By solving Eq.~(\ref{eq:in potential}), we can get the physical mass that includes the contribution of both the hadron $BC$ channel and the bare state. The two terms inside the parenthesis of Eq. (\ref{eq:in potential}) are responsible for the mass shift.

For obtaining the solution of Eq. (\ref{eq:in potential}), we use a set of complete base expansion method, where the complete basis can be chosen as the Harmonic oscillator basis, the Gaussian basis, and so on. For example, we use the Gaussian basis to replace the $\chi_{BC}(\textbf{p})$ in the Eq. (\ref{eq:in potential})~\cite{Hiyama:2003cu,Hiyama:2012sma}
\begin{eqnarray}
	\chi_{BC}({\bf p})=\sum_{i=1}^{N_{max}}C_{il}\phi_{ilm}^p({\bf p}),
\end{eqnarray}
where $C_{il}$ is the coefficient of the corresponding basis, and $\phi_{ilm}^p({\bf p})$ is the Gaussian basis. In the coordinate space,
\begin{eqnarray}\label{eq:Gr}
\phi_{nlm}^r(\nu_n,\textbf{r})=N_{nl}r^le^{-v_nr^2}Y_{lm}(\hat{\textbf{r}}),
\end{eqnarray}
where $N_{nl}$ is normalization constant. By the Fourier transform, the Gaussian basis in momentum space can be written as
\begin{eqnarray}
\phi_{nlm}^p(\nu_n,\textbf{p})=(-i)^l\phi_{nlm}^r\left(\frac{1}{4\nu_n},\textbf{p}\right).\label{eq:Gp}
\end{eqnarray}
In Eqs.~(\ref{eq:Gr}) and (\ref{eq:Gp}), $v_n$ is the Gaussian ranges, i.e.,
\begin{eqnarray}
	v_n=1/r_n^2,\quad r_n=r_1a^{n-1}\quad (n=1,2\dots N_{max}).
\end{eqnarray}

With the above Gaussian basis, all the Hamilton matrix elements can be expressed in simple forms. We define
\begin{equation}\label{eq:matrixelement}
\begin{split}
T_{fi}=&\int \textrm{d}^3\textbf{p}^\prime \phi_{flm}^{p*}(\nu_f,\textbf{p}^{\prime})E_{BC}(\textbf{p}^{\prime})\phi_{ilm}^p(\nu_i,\textbf{p}^{\prime}),\\
{\cal M}_{fi}=&\int \textrm{d}^3\textbf{p}^{\prime}\textrm{d}^3\textbf{p}\frac{H_{\Psi_0\to BC}^*(\textbf{p})H_{\Psi_0\to BC}(\textbf{p}^{\prime})}{M-M_0}\\
&\times\phi_{flm}^{p*}(\nu_f,\textbf{p}^{\prime})\phi_{ilm}^p(\nu_i,\textbf{p}),\\
V_{fi}=&\int \textrm{d}^3\textbf{p}^{\prime}\textrm{d}^3\textbf{p} \phi_{flm}^{p*}(\nu_f,\textbf{p}^{\prime})V_{BC\to BC}(\textbf{p},\textbf{p}^{\prime})\phi_{ilm}^p(\nu_i,\textbf{p}),\\
N_{if}=&\int \textrm{d}^3 \textbf{r}^{\prime}\phi_{flm}^{r*}(\nu_f,\textbf{r}^{\prime})\phi_{ilm}^r(\nu_i,\textbf{r}^{\prime}).
\end{split}
\end{equation}
With the above matrix elements, Eq. (\ref{eq:in potential}) is equivalent to following general eigenvalue equation
\begin{eqnarray}\label{eq:EigM}
	\sum_{i=1}^{N_{max}}C_{il}(T_{fi}+{\cal M}_{fi}+V_{fi})=M\sum_{i=1}^{N_{max}}C_{il}N_{fi}.
\end{eqnarray}
Here the coefficient $C_{il}$ can be solved by the Rayleigh-Ritz variational principle. Because both sides of Eq. (\ref{eq:EigM}) depend on $M$, we are dealing with a special eigenvalue equation. First we replace $M$ with $M_E$ on the right side. Then we scan all possible $M$ on the left side of the new equation in a reasonable range and obtain the eigenvalues $M_E$. At last, the solution comes as $M$ equals to $M_E$.

If the interaction $V_{BC\to BC}({\bf p},{\bf p}^\prime)$ of the direct hadron-hadron channel is neglected, we can extract the approximate mass $M$
\begin{eqnarray}\label{eq:outpotential}
	M=M_0+\int \frac{|H_{A^{\rm bare}\to BC}(\textbf{p})|^2}{M-E_{BC}(\textbf{p})}\textrm{d}^3\textbf{p},
\end{eqnarray}
and obtain approximate wave function $\chi_{BC}({\bf p})$
\begin{equation}
\chi_{BC}({\bf p})=\frac{H_{\Psi_0\to BC}({\bf p})}{M-E_{BC}({\bf p})}c_0.
\end{equation}

\section{Detailed interactions}\label{sec3}
In this section, we provide the $D^*N$ interaction in Sec.~\ref{sec:DNint} and the coupling between the bare state and the $D^*N$ in Sec.~\ref{sec:DNbareCoup}. These determine the coupled-channel effects of $\Lambda_c(2940)^+$ and $\Lambda_c(2910)^+$.

\subsection{$D^*N$ INTERACTION}\label{sec:DNint}
Firstly, we focus on the detailed potential $V_{BC\to BC}$ for the $S$-wave $D^*N$ interaction, and we employ the chiral effective field theory, which is a powerful instrument to study the hadron-hadron interaction~\cite{Kang:2013uia,Weinberg:1978kz,Gasser:1983yg,Scherer:2012xha,Meng:2022ozq,Machleidt:2020vzm,Meissner:2015wva,Meissner:2014lgi,Epelbaum:2008ga,Bernard:1995dp}. In the heavy flavor hadron systems, the application of the chiral effective field theory has led to some achievements for predicting the bound states of $\bar{B}^{(*)}\bar{B}^{(*)}$, $DD^*$, $D\bar{D}^*$, $\Sigma_c\bar{D}^{(*)}$, and so on~\cite{Liu:2012vd,Wang:2018atz,Xu:2017tsr,Xu:2021vsi,Meng:2019ilv,Meng:2019dba,Meng:2019nzy,Wang:2019ato,Chen:2022iil,Meng:2022ozq,Wang:2019nvm}. With the experience in these works, the $D^*N$ interaction can also be studied within the chiral effective field theory.

We will investigate four channels in the $D^*N$ system, i.e., isospin $I=0$, $1$ and spin $J=1/2$, $3/2$. Therefore, we consider the following flavor wave functions
\begin{eqnarray*}
    &&\left|1,1\right\rangle=\left|pD^{*+}\right\rangle,\\
    &&\left|1,0\right\rangle=\frac{1}{\sqrt{2}}\left(\left|pD^{*0}\right\rangle-\left|nD^{*+}\right\rangle\right),\\
    &&\left|1,-1\right\rangle=\left|nD^{*0}\right\rangle,\\
    &&\left|0,0\right\rangle=\frac{1}{\sqrt{2}}\left(\left|pD^{*0}\right\rangle+\left|nD^{*+}\right\rangle\right).
\end{eqnarray*}
In the following, we first give the chiral Lagrangians and Feynman diagrams including tree and one-loop diagrams of the $D^*N$ system. Then, we provide the $D^*N$ effective potentials in momentum space at the next-to-leading order $\mathcal{O}(\epsilon^2)$. They include the contact term, one-pion-exchange, and two-pion-exchange contributions, which approximatively corresponds to the short-range, long-range, and middle-range interactions, respectively. In addition, we also consider the $\Delta(1232)$ contribution. Through the Fourier transformation, we also obtain the effective potentials in coordinate space.

\begin{figure*}[!htbp]
	\centering
	\includegraphics[width=0.95\linewidth]{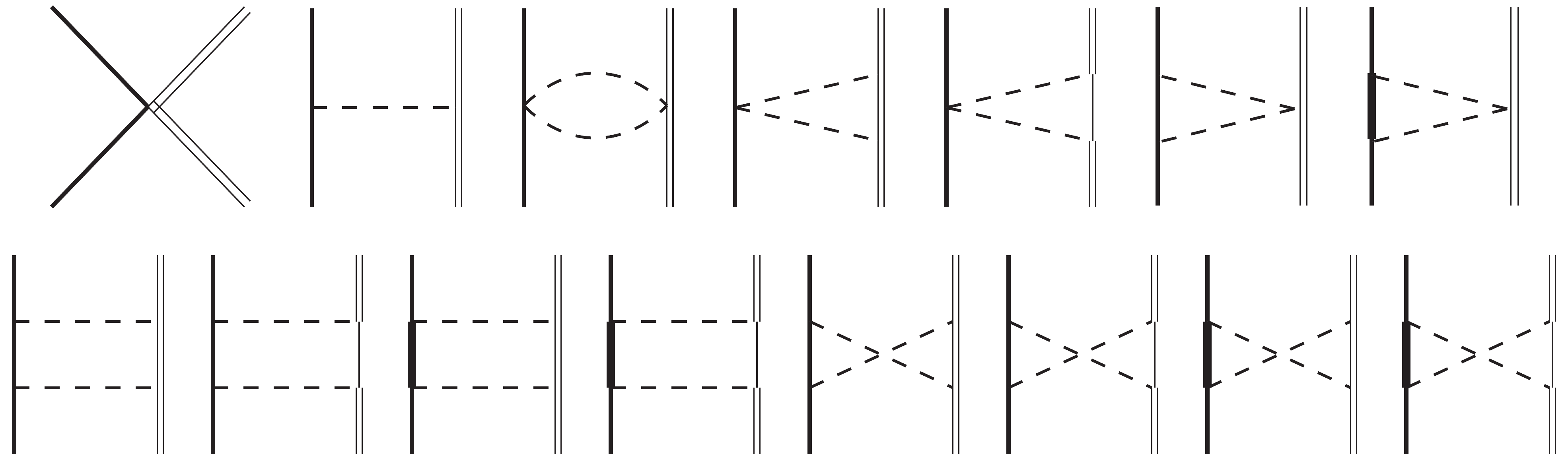}
	\caption{The Feynman diagrams of contact term at $\mathcal{O}(\epsilon^0)$, one-pion-exchange at $\mathcal{O}(\epsilon^0)$, and two-pion-exchange at $\mathcal{O}(\epsilon^2)$. The thin, double-thin, thick, heavy-thick, and dashed lines denote the $D$, $D^*$, $N$, $\Delta(1232)$, and pion field, respectively.}\label{fig:feynman}
\end{figure*}

The leading order $\pi N$ Lagrangian~\cite{Weinberg:1990rz,Weinberg:1991um,Bernard:1992qa} is given by
\begin{eqnarray}
\mathcal{L}_{\mathcal{N}\varphi}=\bar{\mathcal{N}}(iv\cdot
\mathcal{D}+2g_a\mathcal{S}\cdot u)\mathcal{N},
\end{eqnarray}
where $\mathcal{N}=(p,n)^T$ denotes the the large component of the nucleon field under the nonrelativistic reduction, $v=(1,0,0,0)$ stands for the four-velocity of the nucleon, $\mathcal{D}_\mu=\partial_{\mu}+\Gamma_{\mu}$ is covariant derivative, $g_a$ is the axial-vector coupling constant, and $\mathcal{S}^\mu=\frac{i}{2}\gamma_5\sigma^{\mu\nu}v_{\nu}$ denotes the Pauli-Lubanski spin vector. The chiral connection $\Gamma_{\mu}$ and axial-vector current $u_{\mu}$ are expressed as
\begin{eqnarray}
\Gamma_\mu&\equiv&\frac{1}{2}\left[\xi^\dag,\partial_\mu\xi\right]\equiv\tau^i\Gamma_\mu^i,\quad u_\mu\equiv\frac{i}{2}\left\{\xi^\dag,\partial_\mu\xi\right\}\equiv\tau^i\omega_\mu^i,\quad\quad
\end{eqnarray}
where $\tau^i$ is  a $2$-component Pauli matrix in the isospin space, i.e.,
\begin{eqnarray}
\xi^2=U=\exp\left(\frac{i\phi}{f_\pi}\right),\qquad
\phi=\left(\begin{array}{cc}
\pi^0&\sqrt{2}\pi^+\\
\sqrt{2}\pi^-&-\pi^0
\end{array} \right),
\end{eqnarray}
and $f_{\pi}$ is the pion decay constant.

The $D^{(*)}\pi$ Lagrangian at the leading order is given by~\cite{Wise:1992hn,Burdman:1992gh,Yan:1992gz,Manohar:2000dt}
\begin{eqnarray}\label{Meson_Lag_SF}
\mathcal{L}_{\mathcal{H}\varphi}=i\langle \mathcal{H}v\cdot\mathcal{D}\bar{\mathcal{H}}\rangle-\frac{1}{8}\delta_b\langle\mathcal{H}\sigma^{\mu \nu}\bar{\mathcal{H}}\sigma_{\mu \nu}\rangle +g\langle\mathcal{H}\slashed{u}\gamma_{5}\bar{\mathcal{H}}\rangle,\qquad
\end{eqnarray}
where $\langle\cdots\rangle$ denotes the trace in the spinor space, $\delta_b$ is mass
shift between $D^*$ and $D$, i.e., $\delta_b=m_{D^\ast}-m_{D}$, and it does not disappear in the chiral limit. $g$ stands for the axial coupling constant. The $\mathcal{H}$ represents the super-field for the charmed mesons, which reads
\begin{eqnarray}
\mathcal{H}&=&\frac{1+\slashed{v}}{2}\left(P_\mu^\ast\gamma^\mu+iP\gamma_5\right),\nonumber\\
\bar{\mathcal{H}}&=&\gamma^0H^\dag\gamma^0=\left(P_\mu^{\ast\dag}\gamma^\mu+iP^\dag\gamma_5\right)\frac{1+\slashed{v}}{2}.
\end{eqnarray}
with $P=(D^0,D^+)^T$ and $P^\ast=(D^{\ast0},D^{\ast+})^T$, respectively. 

The leading order contact Lagrangian describes the short distance interaction between the nucleon and charmed meson~\cite{Wang:2020dhf}, which can be written as
\begin{eqnarray}\label{Lag_NH}
\mathcal{L}_{\mathcal{NH}}&=&D_a\bar{\mathcal{N}}\mathcal{N}\langle\bar{\mathcal{H}}\mathcal{H}\rangle+D_b\bar{\mathcal{N}}\gamma_\mu\gamma_5\mathcal{N}\langle\bar{\mathcal{H}}\gamma^\mu\gamma_5\mathcal{H}\rangle\nonumber\\
&&+E_a\bar{\mathcal{N}}\tau_i\mathcal{N}\langle\bar{\mathcal{H}}\tau_i\mathcal{H}\rangle+E_b\bar{\mathcal{N}}\gamma_\mu\gamma_5\tau_i\mathcal{N}\langle\bar{\mathcal{H}}\gamma^\mu\gamma_5\tau_i\mathcal{H}\rangle,\nonumber\\
\end{eqnarray}
where $D_a$, $D_b$, $E_a$ and $E_b$ are four low energy constants (LECs). 

Finally, considering the strong coupling between $\Delta(1232)$ and $\pi N$~\cite{Machleidt:1987hj,Holinde:1977rh,Krebs:2007rh,Epelbaum:2007sq,Kaiser:1998wa}, the Lagrangian of the $\Delta$-$N$-$\pi$ coupling~\cite{Hemmert:1997ye} is given by
\begin{eqnarray}
\mathcal{L}_{\Delta\varphi}&=&-\bar{\mathcal{T}}_i^\mu(iv\cdot\mathcal{D}^{ij}-\delta^{ij}\delta_a+2g_1\mathcal{S}\cdot u^{ij})g_{\mu\nu}\mathcal{T}_j^\nu,\quad\\
\mathcal{L}_{\Delta\mathcal{N}\varphi}&=&2g_\delta(\bar{\mathcal{T}}_i^\mu
g_{\mu\alpha}\omega_i^\alpha\mathcal{N}+\bar{\mathcal{N}}\omega_i^{\alpha\dagger}g_{\alpha\mu}\mathcal{T}_i^\mu),
\end{eqnarray}
where $\delta_a=m_\Delta-m_N$ and $g_1=\frac{9}{5}g_a$ \cite{Hemmert:1997ye}. $g_\delta$ is the coupling constant for $\Delta N\pi$ vertex. The matrix form of $\mathcal{T}_i^\mu$ reads
\begin{eqnarray}
\mathcal{T}_\mu^1&=&\frac{1}{\sqrt{2}}\left(\begin{array}{c}\Delta^{++}-\frac{1}{\sqrt{3}}\Delta^0\\\frac{1}{\sqrt{3}}\Delta^+-\Delta^-\end{array} \right)_\mu,\nonumber\\
\mathcal{T}_\mu^2&=&\frac{i}{\sqrt{2}}\left(\begin{array}{c}\Delta^{++}+\frac{1}{\sqrt{3}}\Delta^0\\\frac{1}{\sqrt{3}}\Delta^++\Delta^-
\end{array} \right)_\mu,\nonumber\\
\mathcal{T}_\mu^3&=&-\sqrt{\frac{2}{3}}\left(\begin{array}{c}\Delta^{+}\\
\Delta^0
\end{array} \right)_\mu.
\end{eqnarray}
Here, $\mathcal{T}_i^\mu$ denotes the spin-$3/2$ and isospin-$3/2$ field $\Delta(1232)$ in the non-relativistic reduction.

In the framework of the heavy hadron chiral perturbation theory, the scattering amplitudes of the $D^{\ast}N$ system can be expanded order by order in powers of a small quantity $\varepsilon=q/\Lambda_\chi \sim 1$ GeV, where $q$ is either the momentum of Goldstone bosons or the residual momentum of heavy flavor hadrons, and $\Lambda_\chi$ represents either the chiral breaking scale or the mass of a heavy hadron. The expansion respects the power counting rule \cite{Weinberg:1990rz,Weinberg:1991um}. The Feynman diagrams of $\mathcal{O}(\epsilon^0)$ contact term, $\mathcal{O}(\epsilon^0)$ one-pion-exchange, and $\mathcal{O}(\epsilon^2)$ two-pion-exchange are illustrated in Fig. \ref{fig:feynman}. 

With the chiral Lagrangians and these Feynman diagrams, we can obtain their Feynman amplitudes $\mathcal{M}$. Then we use the Breit approximation $\mathcal{V}=-\mathcal{M}/(\Pi_i 2m_i \Pi_f 2m_f)^{1/2}$ to relate the scattering amplitude $\mathcal{M}$ to the effective potential $\mathcal{V}$~\cite{Wang:2020dhf}, where $m_i$ and $m_f$ are the masses of the initial and final states, respectively. These effective potentials of the Feynman diagrams in Fig. \ref{fig:feynman} consist of following parts
 \begin{eqnarray}
 	\mathcal{V}_{\rm total}=\mathcal{V}_{\rm{contact}}^{\rm{LO}}+\mathcal{V}_{1-\pi}^{\rm{LO}}+\mathcal{V}_{2-\pi}^{\rm{NLO}},
 \end{eqnarray}
where $\mathcal{V}_{\rm{contact}}^{\rm{LO}}$, $\mathcal{V}_{1-\pi}^{\rm{LO}}$, and $\mathcal{V}_{2-\pi}^{\rm{NLO}}$ denote $\mathcal{O}(\epsilon^0)$ contact term, $\mathcal{O}(\epsilon^0)$ one-pion-exchange, and $\mathcal{O}(\epsilon^2)$ two-pion-exchange potentials, respectively. 

The $\mathcal{V}_{2-\pi}^{\rm{NLO}}$ is the sum of the football diagram, triangle diagram, box diagram, and crossed box diagram potentials in Fig.~\ref{fig:feynman}, i.e., $\mathcal{V}_{2-\pi}^{\rm{NLO}}=\mathcal{V}_{\rm football}+\mathcal{V}_{\rm triangle}+\mathcal{V}_{\rm box}+\mathcal{V}_{\rm crossed-box}$. These two-pion-exchange diagrams need some loop functions $J_{ij}^F$, $J_{ij}^T$, $J_{ij}^B$, and $J_{ij}^R$, which can be found in Refs.~\cite{Wang:2019ato,Meng:2019ilv,Liu:2012vd,Wang:2018atz,Xu:2017tsr}. In order to obtain the effective potentials, the two-particle-reducible (2PR) contribution should be subtracted from these crossed box diagrams by the principal value integral method in the Appendix~B of Ref.~\cite{Wang:2019ato}. Additionally, the divergent parts in two-pion-exchange diagrams can be absorbed by unrenormalized LECs in Ref.~\cite{Wang:2020dhf}. In this work, all the parameters for the $D^*N$ interaction are from Ref.~\cite{Wang:2020dhf}. 

\begin{figure*}[htbp]
	\centering
	\includegraphics[width=0.85\linewidth]{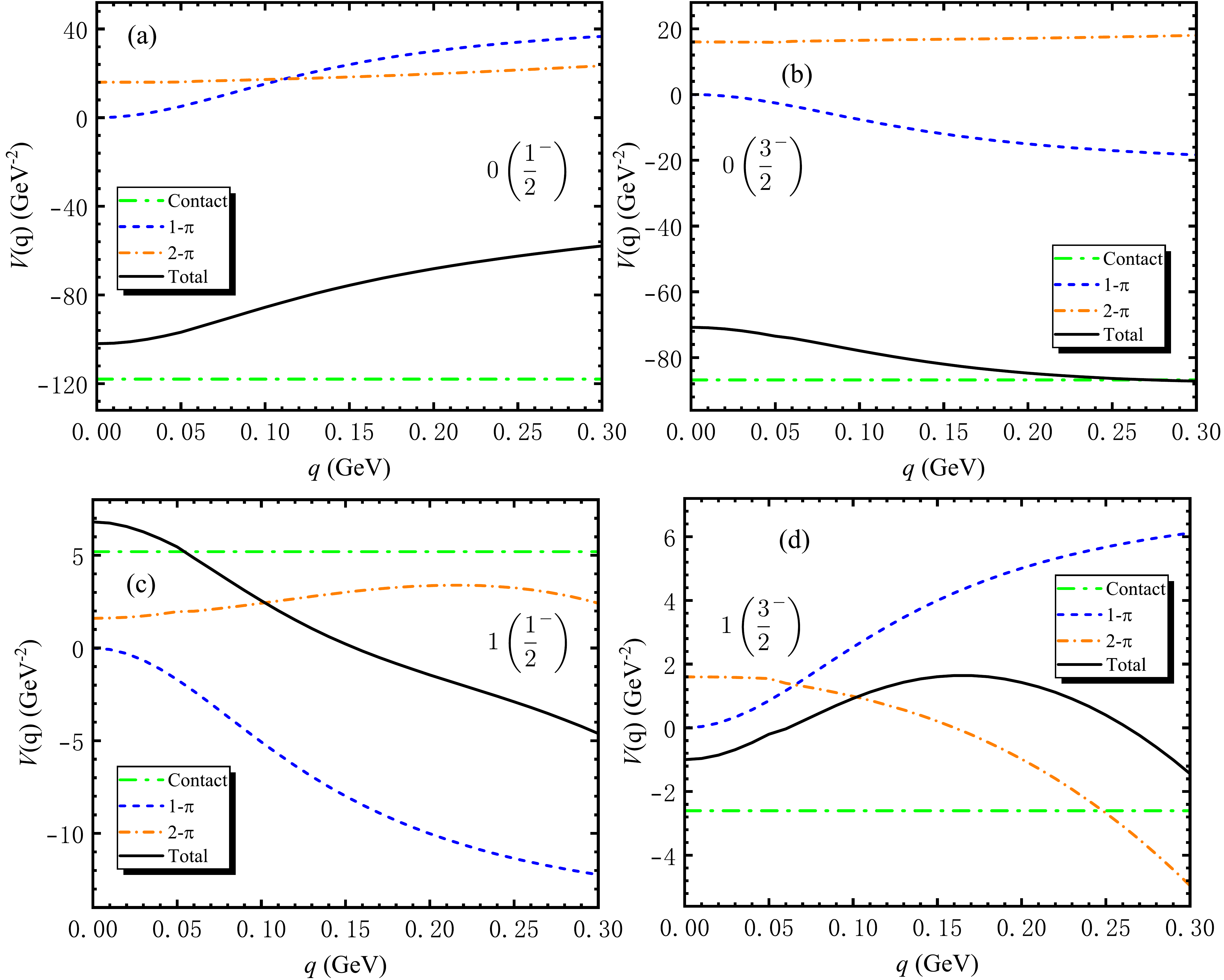}
	\caption{(color online) The $D^*N$ interaction potential in momentum space with the cutoff parameter $\Lambda=0.4$ GeV, and $q=|\bf q|$ is the transfer momentum between $D^*$ and $N$. Here, the green dot, blue dot, and red dot lines describe the contact term potential at $\mathcal{O}(\epsilon^0)$, one-pion-exchange potential at $\mathcal{O}(\epsilon^0)$, and two-pion-exchange potential at $\mathcal{O}(\epsilon^2)$, respectively, while the black solid line denotes the total potential.}\label{fig:potential}
\end{figure*}

In Fig.~\ref{fig:potential}, we give the $D^*N$ effective potential in momentum space, which can help us to understand the $D^*N$ interaction clearer. 
For the $\left[D^*N\right]_{J=1/2}^{I=0}$ channel in Fig.~\ref{fig:potential} (a), the $\mathcal{O}(\epsilon^0)$ one-pion-exchange potential and $\mathcal{O}(\epsilon^2)$ two-pion-exchange potential are both repulsive. But their repulsive interaction is rather weak. The attractive interaction is dominantly provided by the $\mathcal{O}(\epsilon^0)$ contact interaction.
In Fig.~\ref{fig:potential}~(b), the $\mathcal{O}(\epsilon^0)$ one-pion-exchange potential is attractive and the $\mathcal{O}(\epsilon^2)$ two-pion-exchange potential is repulsive, but both of them are rather weak. However, the potential of $\mathcal{O}(\epsilon^0)$ contact term has strong attractive interaction. 
For the $\left[D^*N\right]_{J=1/2}^{I=1}$ channel in Fig.~\ref{fig:potential}~(c), the potential of $\mathcal{O}(\epsilon^0)$ contact term and $\mathcal{O}(\epsilon^2)$ two-pion-exchange are both repulsive, but the $\mathcal{O}(\epsilon^0)$ one-pion-exchange is attractive. The total potential has rather weak attraction.
From Fig.~\ref{fig:potential} (d), the total potential are not attractive enough. Therefore, the two isoscalar channels $I(J^P)=0(1/2^-)$ and $I(J^P)=0(3/2^-)$ provide the stronger attractive interaction in momentum space, and we identify the attraction in two $I=1$ channels is weak. In the following parts, we mainly consider the two isospin $I=0$ channels. Furthermore, the $\Delta(1232)$ plays a particular role in the $D^*N$ interaction because the coupling between $\Delta(1232)$ and $\pi N$ system is very strong. The total potential is sensitive to $\Delta(1232)$, which determines its importance in the $D^*N$ interaction.

Based on the obtained effective potential in momentum space $\mathcal{V}({\bf q})$, the effective potential in coordinate space $\mathcal{V}({\bf r})$ can be obtained by the following Fourier transformation
\begin{eqnarray}\label{eq:FT}
	V_{BC\to BC}({\bf r})=\int\frac{\textrm{d}^3{\bf q}}{(2\pi)^3}e^{i{\bf q}\cdot\textbf{r}}\mathcal{V}({\bf q})\mathcal{F}({\bf q}).
\end{eqnarray}
Here, $\mathcal{F}(\textbf{q})={\rm e}^{-\textbf{q}^{2n}/\Lambda^{2n}}$ is the form factor with Gauss form to suppress the high momentum and renormalize the potential~\cite{Wang:2020dhf,Xu:2017tsr}. 

When we employ the chiral effective field theory and use some approximation, such as the transferred energy $q^0$ between $D^*$ and $N$, and the residual energies of $N$ and $D^*$ are all set to zero, then the effective potential of $D^\ast N$ will have the simple form of $V_{BC\to BC}(\textbf{p}^\prime-\textbf{p})$. Thus, the $V_{BC\to BC}({\textbf{p}^\prime}, \textbf{p})$ in Eq. (\ref{eq:expandHBC}) can be defined as

\begin{equation}\nonumber
V_{BC\to BC}({\bf p}^\prime,{\bf p})=\frac{1}{(2\pi)^3}\int V_{BC\to BC}({\bf r}){\rm e}^{i({\bf p}^\prime-{\bf p})\cdot{\bf r}}{\rm d}^3{\bf r}.
\end{equation}

In this scheme, the matrix element $V_{fi}$ in Eq.~(\ref{eq:matrixelement}) can be conveniently calculated by
\begin{equation}
V_{fi}=\int \textrm{d}^3\textbf{r} \phi_{flm}^{r*}(\nu_f,\textbf{r})V_{BC\to BC}(\textbf{r})\phi_{ilm}^r(\nu_i,\textbf{r}).
\end{equation}

\subsection{Interaction between the bare state and $D^*N$ channel}\label{sec:DNbareCoup}
For the interaction Hamiltonian $\hat{H}_I$ between the bare state and $D^*N$ in Eq.~(\ref{eq:1}), we employ the quark-pair-creation model \cite{Micu:1968mk,LeYaouanc:1972vsx}, which has the expression
\begin{eqnarray}
	\hat{H}_I&=&g\int \textrm{d}^3x\bar{\psi}(x)\psi(x).
\end{eqnarray}
Here, $g=2m_q\gamma$, $m_q$ is the mass of the creation quark, and the dimensionless parameter $\gamma$ describes the strength of the quark and antiquark pair creation from the vacuum, which can be determined phenomenologically by the  Okubo-Zweig-Iizuka (OZI) allowed decay widths of charmonia. 

In the non-relativistic limit, $\hat{H}_I$ is equivalent to~\cite{Chen:2016iyi}
\begin{eqnarray}
	\hat{H}_I=&-3\gamma\sum_{m}\left<1,m;1,-m|0,0\right>\int \textrm{d}^3\textbf{p}_{\mu}d^3\textbf{p}_{\nu}\delta\left(\textbf{p}_{\mu}+\textbf{p}_{\nu}\right)\nonumber\\&\times\mathcal{Y}_1^m\left(\frac{\textbf{p}_{\mu}-\textbf{p}_{\nu}}{2}\right)\omega^{\left(\mu,\nu\right)}\phi^{\left(\mu,\nu\right)}\chi_{-m}^{\left(\mu,\nu\right)}b_{\mu}^{\dagger}\left(\textbf{p}_{\mu}\right)d_{\nu}^{\dagger}\left(\textbf{p}_{\nu}\right).\quad\quad
\end{eqnarray}
Here, $\omega$, $\phi$, $\chi$, and $\mathcal{Y}_1^m$ 
describe the color, flavor, spin, and orbital angular momentum functions of the quark pair, respectively. $b_{\mu}^{\dagger}$ and $d_{\nu}^{\dagger}$ are quark and antiquark creation operators, respectively. In this work, we fit the dimensionless parameter $\gamma$ as $9.45$ from the total decay width of $\Sigma_c(2520)$ \cite{ParticleDataGroup:2022pth}.

The masses of bare $\Lambda_c(2P,1/2^-)$ and $\Lambda_c(2P,3/2^-)$ can be obtained by the traditional quark potential models combined with the Gaussian expansion method by solving Eqs.~(\ref{eq:3}) and (\ref{eq:h0}). In this work, we use different bare masses as input and collect them in Table~\ref{table:bare mass}. In addition, $H_{\Psi_0\to BC}^{*}(\textbf{p})$ in Eq. (\ref{eq:in potential}) can be obtained by calculating the transition amplitude $H_{\Lambda_c^{\rm bare}(2P, 1/2^-)\to D^*N}(\textbf{p})=\langle D^*N,\textbf{p}|\hat{H}_I|\Lambda_c^{\rm bare}(2P,1/2^-)\rangle$ and $H_{\Lambda_c^{\rm bare}(2P,3/2^-)\to D^*N}(\textbf{p})=\langle D^*N,\textbf{p}|\hat{H}_I|\Lambda_c^{\rm bare}(2P,3/2^-)\rangle$ with the quark-pair-creation model. 

\renewcommand\tabcolsep{0.17cm}
\renewcommand\arraystretch{1.4}
\begin{table}[!htbp]
	\caption{The mass of bare $udc$ core in $\Lambda_c(2P,1/2^-)$ and $\Lambda_c(2P,3/2^-)$ in the different potential models. The masses are in units of MeV.}\label{table:bare mass}
	\centering
	\begin{tabular}{c|ccccc}\toprule[1.0pt]\toprule[1.0pt]
		$J^P(nL)$&Ref.~\cite{Chen:2014nyo}&Ref.~\cite{Chen:2016iyi}&Ref.~\cite{Ebert:2011kk}&Ref.~\cite{Luo:2019qkm}&Ref.~\cite{Capstick:1986ter}\\\midrule[1.0pt]
		$1/2^-(2P)$&2989&2980&2983&2996&3030\\
		$3/2^-(2P)$&3000&3004&3005&3012&3035\\
		\bottomrule[1.0pt]	\bottomrule[1.0pt]
	\end{tabular}
\end{table}

\section{RESULTS AND DISCUSSIONS}\label{sec4}
Using the formalism described in Sec.~\ref{sec2} and the detailed interaction in Sec. \ref{sec3}, we can now study the dynamical coupling between the $S$-wave $D^*N$ channel and bare charmed baryon core $\Lambda_c(2P)$. In the quenched quark model, the charmed baryons are simply treated as the three-quark $udc$ baryons~\cite{Chen:2014nyo,Chen:2016iyi,Ebert:2011kk,Luo:2019qkm,Capstick:1986ter}. In the unquenched picture, the physical states $\Lambda_c(2940)^+$ and $\Lambda_c(2910)^+$ consist of both $udc$ core and the $S$-wave $D^*N$ component. In the former works, the authors of Ref.~\cite {Luo:2019qkm} considered the coupling between $udc$ and $D^*N$ but did not involve the $D^*N$ interaction, while this work contains them both. We list the results of the three approaches in Table~\ref{table:RESULT}.

\renewcommand\tabcolsep{0.40cm}
\renewcommand\arraystretch{1.8}
\begin{table*}[!htbp]
	\caption{Comparison among the results from quenched quark model~\cite{Chen:2014nyo,Chen:2016iyi,Ebert:2011kk,Luo:2019qkm,Capstick:1986ter}, unquenched picture without $D^*N$ interaction~\cite {Luo:2019qkm}, and unquenched picture with $D^*N$ interaction in this work. Here, $r_{\rm RMS}$ refers to the root-mean-square radius of the $D^*N$ component, and $P(udc)=c_0^2$ represents the probability of bare $udc$ core in the bound state. }\label{table:RESULT}
	\centering
	\begin{tabular}{c|c|c|ccc|ccc}\toprule[1.0pt]\toprule[1.0pt]
		Cases&\multicolumn{2}{c|}{Quenched picture} &\multicolumn{3}{c|}{Unquenched picture without $D^*N$ interaction} &\multicolumn{3}{c}{Unquenched picture with $D^*N$ interaction}\\\midrule[1.pt]
		$J^P$&\multicolumn{2}{c|}{$M_0$ (MeV)}&$M$ (MeV)&$r_{\rm RMS}$ (fm)&$P(udc)$ $(\%)$&$M$ (MeV)&$r_{\rm RMS}$ (fm)&$P(udc)$ $(\%)$\\\midrule[1.pt]
		$1/2^-$&\multirow{2}{*}{Ref.~\cite{Chen:2014nyo}}&2989&2974&$\times$&$\times$&2936&1.93&16.2\\
		$3/2^-$&&3000&2933&1.67&39.7&2908&1.31&29.4\\\midrule[1.pt]
		$1/2^-$&\multirow{2}{*}{Ref.~\cite{Chen:2016iyi}}&2980&2955&$\times$&$\times$&2934&1.83&21.9\\\
		$3/2^-$&&3004&2935&1.74&37.0&2909&1.31&27.9\\\midrule[1.pt]
		$1/2^-$&\multirow{2}{*}{Ref.~\cite{Ebert:2011kk}}&2983&2962&$\times$&$\times$&2935&1.87&19.8\\
		$3/2^-$&&3005&2935&1.76&36.3&2909&1.32&27.5\\\midrule[1.pt]
		$1/2^-$&\multirow{2}{*}{Ref.~\cite{Luo:2019qkm}}&2996&2985&$\times$&$\times$&2937&2.00&13.4\\
		$3/2^-$&&3012&2937&1.95&31.4&2911&1.33&25.2\\\midrule[1.pt]
		$1/2^-$&\multirow{2}{*}{Ref.~\cite{Capstick:1986ter}}&3030&3036&$\times$&$\times$&2940&2.32&5.08\\
		$3/2^-$&&3035&2943&2.93&15.8&2916&1.38&18.7\\
		\bottomrule[1.0pt]	\bottomrule[1.0pt]
	\end{tabular}
\end{table*}

From the left column in Table~\ref{table:RESULT}, the masses of the $\Lambda_c(2P,1/2^-)$ and $\Lambda_c(2P,3/2^-)$ from quenched quark model are much larger than those of $\Lambda_c(2940)^+$ and $\Lambda_c(2910)^+$.  From the middle columns in Table \ref{table:RESULT}, the theoretical masses decrease due to the coupling between the $udc$ core and $D^*N$, but neither can reach the mass of  lower state $\Lambda_c(2910)^+$~\cite{Luo:2019qkm}. Among them, we can see the coupled-channel effects between the $\Lambda_c(2P,1/2^-)$ and $S$-wave $\left[D^*N\right]_{J=1/2}^{I=0}$ channel is relatively weak, and thus the mass is only suppressed by about 20~MeV and is still above the $D^*N$ threshold. Neither $\Lambda_c(2940)^+$ nor $\Lambda_c(2910)^+$ can match to it. Meanwhile, the coupled effects between $\Lambda_c(2P,3/2^-)$ and $S$-wave $[D^*N]_{J=3/2}^{I=0}$ channel is strong, the mass decreases by about $70$~MeV and locates below $D^*N$ threshold.

From the right columns in Table \ref{table:RESULT}, the $D^*N$ interaction further lowers the theoretical masses of the $\Lambda_c(2P,1/2^-)$ and $\Lambda_c(2P,3/2^-)$ which now become very close to those of $\Lambda_c(2940)^+$ and $\Lambda_c(2910)^+$ respectively.
Thus we assign $\Lambda_c(2910)^+$ with $J^P=3/2^-$ and $\Lambda_c(2940)^+$ with $J^P=1/2^-$. As we can see, the attractive $D^*N$ interaction plays a crucial role to reproduce the experimental $\Lambda_c(2940)^+$ and $\Lambda_c(2910)^+$. 

In the unquenched pictures with or without $D^*N$ interaction, the mass for $J^P=1/2^-$ is larger than that for $3/2^-$, which is different from the quenched case. Such mass-inversion phenomenon also happens in the $N(1535)$-$N(1440)$ case \cite{Liu:2015ktc,Liu:2016uzk}. The mass inversion phenomenon comes from the two factors in this work: (1) under the chiral effective field theory~\cite{Wang:2020dhf}, the energy level in $S$-wave $\left[D^*N\right]_{J=1/2}^{I=0}$ channel is larger than that in the $\left[D^*N\right]_{J=3/2}^{I=0}$ channel; (2) the coupling between $\Lambda_c(2P,3/2^-)$ and $\left[D^*N\right]_{J=3/2}^{I=0}$  is stronger than that between $\Lambda_c(2P,1/2^-)$ and $\left[D^*N\right]_{J=1/2}^{I=0}$ ~\cite {Luo:2019qkm}. These lead to the mass inversion between $\Lambda_c(2P,1/2^-)$ and $\Lambda_c(2P,3/2^-)$.

\begin{figure*}[!htbp]
  	\centering
  	\includegraphics[width=0.85\linewidth]{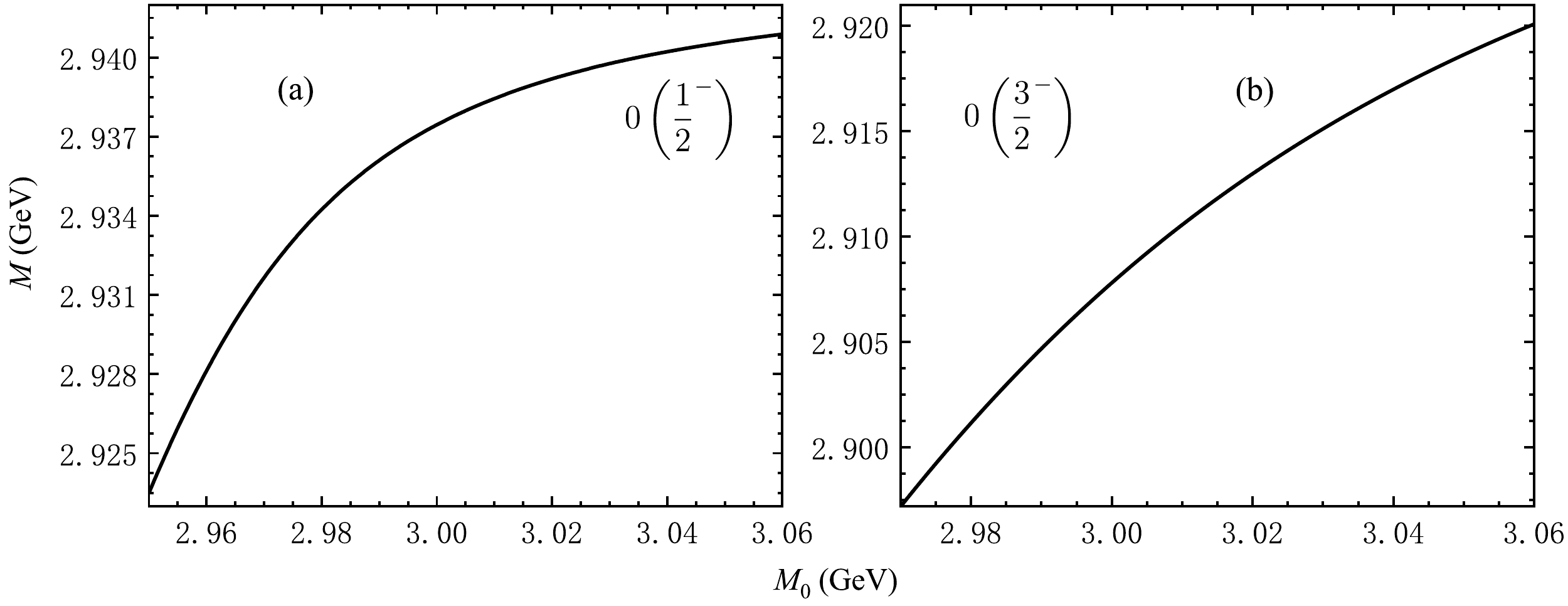}
  	\caption{The physical mass $M$ dependence on the bare mass $M_0$ in $\Lambda_c(2P,1/2^-)$ and $\Lambda_c(2P,3/2^-)$, respectively.}\label{fig:Mdepend}
  \end{figure*}

In Fig. \ref{fig:Mdepend}, we present the physical mass $M$ dependence on the bare mass $M_0$ in $\Lambda_c(2P,1/2^-)$ and $\Lambda_c(2P,3/2^-)$. One can notice that the physical masses increase as the bare masses. There are differences between the vertical axis and horizontal axis values of points in the curves, which shows that the bare $udc$ core has a significant mass shift due to the effects of $S$-wave $D^*N$ channel.
 
   \begin{figure*}[!htbp]
  	\centering
	\includegraphics[width=0.85\linewidth]{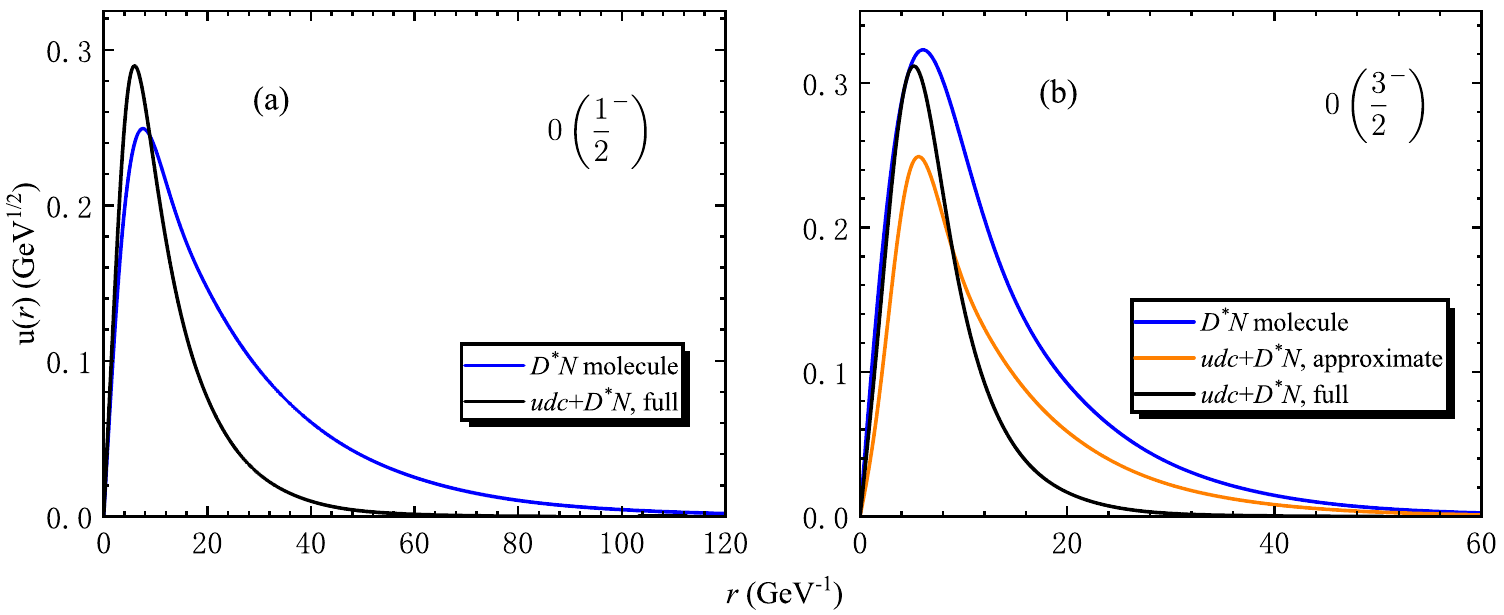}
  	\caption{
  	(color online) The $D^*N$ radial wave functions for the pure $D^* N$ molecule (`$D^* N$ molecule'), $udc$ core dressed by $D^*N$ without considering $D^*N$ interaction ( `$udc+D^*N$, approximate'), and the dressed state with $D^*N$ interaction (`$udc+D^*N$, full'). Here, the cutoff parameter $\Lambda=0.4$ GeV and the bare masses are adopted from Ref.~\cite{Luo:2019qkm}.}\label{M-wavefunction}
  \end{figure*}

By solving Eq. (\ref{eq:in potential}), one can obtain the radial wave function of $D^*N$ channel and the probability amplitude of bare $udc$ core, which can help us to reveal the role of $D^*N$ channel and bare $udc$ core in physical state $\Lambda_c(2940)^+$ and $\Lambda_c(2910)^+$. In Table \ref{table:RESULT}, we also provide the root-mean-square radius of $D^*N$ component and the probability of bare $udc$ core in the bound state, and we can see the $3/2^-$ state is a little thinner and contains more bare $udc$ core than the $1/2^-$ one.

If neglecting the triquark core, the pure $D^*N$ can also form bound states with masses around 2940 MeV with $J^P=1/2^-$ and $3/2^-$~\cite{Wang:2020dhf}. But such assumptions make us difficult to place the bare $2P$ $udc$ states. In Fig.~\ref{M-wavefunction}, we present the $D^*N$ radial wave functions for the three circumstances: the pure $D^* N$ molecule, the $udc$ core dressed by $D^*N$ without considering $D^*N$ interaction, and the dressed state with $D^*N$ interaction. From Fig.~\ref{M-wavefunction}, we can find that the existence of bare state can make the $D^*N$ bind more tightly. These wave functions can be used to analyze other properties of charmed baryons, and it can help us to distinguish which model is better in future.
  
\renewcommand\tabcolsep{0.17cm}
\renewcommand\arraystretch{1.4}
\begin{table}[!htbp]
\caption{The results with $\Delta(1232)$ turned off.}\label{table:delta1232}
\centering
\begin{tabular}{c|cccc}
\toprule[1.0pt]\toprule[1.0pt]
$J^P$ (MeV)&$M_0$ (MeV)&$M$ (MeV)&$r_{\rm RMS}$ (fm)&$P(udc)$ ($\%$)\\\midrule[1.0pt]
$1/2^-$&2996&2930&1.67&12.7\\
$3/2^-$&3012&2902&1.26&22.4\\
\bottomrule[1.0pt]	\bottomrule[1.0pt]
\end{tabular}
\end{table}
  
As mentioned earlier, the $\Delta(1232)$ plays an important role in these two physical states. If ignoring the contribution of $\Delta(1232)$, the two-pion-exchange potential becomes smaller. We list the results in Table \ref{table:delta1232} if turning off $\Delta(1232)$. From the table, the masses of dressed states decrease by about 10~MeV. In two isoscalar states $\left[D^*N\right]_{J=1/2}^{I=0}$ and $\left[D^*N\right]_{J=3/2}^{I=0}$, the total potentials become more attractive, which bring these two states thinner.

How to get rid of the cutoff dependence in nonperturbative calculations is still an outstanding problem both in hadron physics and nuclear physics. In order to investigate the cutoff dependence of our results, we list the charmed baryon masses with two different cutoffs in Table \ref{table:lambda}. One can see that the unquenched masses all become lower with the cutoff becomes larger, and the results at $\Lambda=0.4$ GeV are more in line with the experimental data. Currently, we can only provide reasonable results in a very narrow range of cutoff due to the restriction of the validity region of the chiral effective field theory.

Moreover, $\Lambda_c(2910)^+$ and $\Lambda_c(2940)^+$ have small decay widths~\cite{ParticleDataGroup:2022pth,Belle:2022hnm,BaBar:2006itc,Belle:2006xni,LHCb:2017jym}. In our unquenched calculation, the $\Lambda_c(2910)^+$ and $\Lambda_c(2940)^+$ lie below the $D^*N$ threshold (see Fig.~\ref{fig:1}). Thus, the decays to $D^*N$ channel are kinematically forbidden, which may result in small decay widths of these two states. The strong decay widths of $\Lambda_c(2P,1/2^-)$ and $\Lambda_c(2P,3/2^-)$ have been studied in Ref.~\cite{Lu:2018utx}, and the results are consistent with our assignment, i.e., $\Lambda_c(2910)^+$ with spin-$3/2$ and $\Lambda_c(2940)^+$ with spin-$1/2$.

\renewcommand\tabcolsep{0.7cm}
\renewcommand\arraystretch{1.4}
\begin{table}[!htbp]
	\caption{The cutoff dependence of the charmed baryon masses in units of MeV. The quenched masses are taken from Ref.~\cite{Luo:2019qkm}}\label{table:lambda}
	\centering
	\begin{tabular}{c|c|c}
		\toprule[1.0pt]\toprule[1.0pt]
		$J^P$&$\Lambda=0.4$ GeV&$\Lambda=0.6$ GeV\\\midrule[1.0pt]
		$1/2^-$&2937&2924\\
		$3/2^-$&2911&2869\\
		\bottomrule[1.0pt]\bottomrule[1.0pt]
	\end{tabular}
\end{table}

In our calculation, we gave the possible interpretation that the $J^P$ for $\Lambda_c(2940)$ is $1/2^-$, which is in clear conflict with the preferred $3/2^-$ assignment from the LHCb experiment \cite{LHCb:2017jym}. However, our results cannot be completely ruled out by the current experiment, and some other articles also agree with the $J^P=1/2^-$ assignment \cite{He:2006is,Ebert:2011kk,Wang:2015rda,Xie:2015zga,Huang:2016ygf,Wang:2020dhf,Yu:2022ymb}. Moreover, the LHCb Collaboration concludes that the other solutions with spins 1/2 to 7/2 cannot be excluded \cite{LHCb:2017jym}. 

The $D^0p$ mass region in the amplitude fit is $2.8\sim 3.0$ GeV in Ref. \cite{LHCb:2017jym}, but only $\Lambda_c(2940)$ was included. Two resonances $\Lambda_c(2940)$ and $\Lambda_c(2910)$ should be taken into account at the same time, and the conclusion may be changed from the new fit in experiment. The $J^P$ assignment for $\Lambda_c(2940)$ should further be measured by other way like the partial wave analysis. Our results should also be further checked by analyzing their other properties in theory. 
\section{Summary}\label{sec5}

With the accumulation of experimental data, a series of charmed baryons have been reported in the past decades. However, some cannot be fitted into the charmed baryon family very well. Inspired by the reported $\Lambda_c(2910)^+$ and $\Lambda_c(2940)^+$ \cite{ParticleDataGroup:2022pth,Belle:2022hnm,BaBar:2006itc,Belle:2006xni,LHCb:2017jym}, we use an unquenched picture to study them by considering $S$-wave $D^*N$ channel coupled with the bare $udc$ core ($\Lambda_c(2P)$), which gives a unified description of $\Lambda_c(2910)^+$ and $\Lambda_c(2940)^+$.

We take into account of two important factors in this work, i.e., the $D^*N$ interaction and the triquark-$D^*N$ coupling. In our unquenched picture, we reproduce the masses of $\Lambda_c(2910)^+$ and $\Lambda_c(2940)^+$ and assign $\Lambda_c(2910)^+$ to $J^P=3/2^-$ and $\Lambda_c(2940)^+$ to $1/2^-$. The results show the unquenched effects lead to the mass inversion phenomenon in the two states, and the $D^*N$ channel is important. 

In the present work, we find the $D^*N$ interaction is crucial in forming the physical states. Let us look at Fig.~\ref{fig:1} again. There are large gaps between the quenched quark model results and the experimental masses. The $3/2^-$ mass is pulled down more than the $1/2^-$ one by the coupling between the triquark core and the $D^*N$, which causes the mass inversion phenomenon. The mass spectrum moves down further after considering the attractive $D^*N$ interaction, and eventually consistent with the experiments. 

From our obtained results,  we see $\Lambda_c(2910)^+$ and $\Lambda_c(2940)^+$ contain a significant $D^*N$ component, and the bare state can cause the $D^*N$ binding more compactly. In addition, we also study the influence of $\Delta(1232)$ in the $D^*N$ interaction. If neglecting $\Delta(1232)$, the dressed masses would be about 10~MeV smaller than before.

Revealing the mixed structure of $\Lambda_c(2910)^+$ and $\Lambda_c(2940)^+$ in the unquenched picture, we expect it can be further verified by other theoretical approaches like lattice QCD simulations. In addition to mass, other properties of these two states should also be analyzed within this picture in future. More importantly, we strongly suggest experiment to give a further study in future, which can provide more hints to uncover the nature of $\Lambda_c(2910)^+$ and $\Lambda_c(2940)^+$.

\section*{ACKNOWLEDGMENTS}
This project is supported by the National Natural Science Foundation of China under Grants No. 12175091, No. 11965016, No. 12047501, No. 12105072, No. 12005168, and CAS Interdisciplinary Innovation Team. B.W is also supported by the Youth Funds of Hebei Province (No. A2021201027) and the Start-up Funds for Young Talents of Hebei University (No. 521100221021). H. X. also acknowledges the Natural Science Foundation
of Gansu province under Grant No. 22JR5RA171.

\end{document}